\documentclass[twocolumn,prl,floatfix,superscriptaddress]{revtex4}
\usepackage{graphicx, subfigure, amsfonts,amssymb,amsmath, array, gensymb,fontenc,color, lipsum}
\usepackage{tikz}

\begin{document}

\title{{Atomic-scale defects in the two-dimensional ferromagnet CrI$_3$ from first principles }}

\author{Michele Pizzochero}
\email{michele.pizzochero@epfl.ch}

\affiliation{Institute of Physics, Ecole Polytechnique F\'ed\'erale de Lausanne (EPFL), CH-1015 Lausanne, Switzerland}
\affiliation{National Centre for Computational Design and Discovery of Novel Materials (MARVEL),  Ecole Polytechnique F\'ed\'erale de Lausanne (EPFL), CH-1015 Lausanne, Switzerland }

\date{\today}

\begin{abstract} 
The family of atomically thin magnets holds great promise for a number of prospective applications in magneto-optoelectronics, with CrI$_3$  arguably being its most prototypical member. However, the formation of defects in this system remains unexplored to date. Here, we investigate native point defects in monolayer CrI$_3$ by means of first-principles calculations. We consider a large set of  intrinsic impurities and address their atomic structure, thermodynamic stability, diffusion and aggregation tendencies as well as local magnetic moments. Under thermodynamic equilibrium, the most stable defects are found to be either Cr or I atomic vacancies along with their complexes, depending on the chemical potential conditions.  These defects are predicted to be quite mobile at room temperature and to exhibit a strong tendency to agglomerate. In addition, our calculations indicate that the defect-induced deviation from the nominal stoichiometry largely impacts the local magnetic moments, thereby suggesting a marked interplay between magnetism and disorder in CrI$_3$. Overall, this work portrays a comprehensive picture of intrinsic point defects in monolayer CrI$_3$ from a theoretical perspective.
\end{abstract}

\maketitle

\section{I.\ Introduction}
Atomically thin magnets have recently joined the ever-growing family of two-dimensional materials, and are currently emerging as suitable candidates for novel magneto-optoelectronics devices in the ultimate limit of atomic thickness \cite{Burch2018, Gibertini19, Gong19, Mak2019, Huang2017}.  Among them, ultrathin films of CrI$_3$ have attracted a great deal of interest due to the extensive control that can be achieved over their magnetic properties \cite{Huang2017, Song19, Huang2018, Jiang2018, Klein1218}. In particular, such films have been isolated upon exfoliation from their layered bulk counterpart, and consist of a honeycomb network of Cr$^{3+}$ ions sandwiched between a pair of atomic planes of I ligands \cite{Huang2017, McGuire15}. The intrinsic magnetism of bulk CrI$_3$ is preserved when the crystal is thinned to the few-layer and eventually single-layer regime of thickness, owing to the key role played by the single-ion anisotropy which prevents thermal fluctuations to quench the magnetically ordered phases up to 45 K \cite{Mermin66, Huang2017}. While in ultrathin films of CrI$_3$ both in-plane ferromagnetic and off-plane antiferromagnetic interactions coexist \cite{Huang2017, Sivadas18}, in the single-layer regime the exchange interactions are restricted to the in-plane direction solely, where Cr$^{3+}$ ions act as spin-$3/2$ centers and couple ferromagnetically through I-mediated super-exchange channels \cite{Lado17, Pizzochero19a}.

Under thermodynamic equilibrium conditions, the entropic contribution to the overall free energy of a solid is responsible for the ubiquitous presence of defects at finite temperatures. Although disorder often affects the properties of materials in a detrimental manner, in many cases the incorporation of a certain amount of impurities has proven instrumental in expanding the functionalities of the hosting system. This is especially true in ultrathin crystals, where the reduced dimensionality of the lattice enhances the impact of imperfections \cite{review, susi-2d17}, as it has been demonstrated \emph{e.g.}\ for elemental monolayers like graphene \cite{gra1, gra2, gra3, gra4, gra5}, silicene \cite{sil1, sil2, sil3} or phosopherene \cite{phos1, phos2}, as well as transition metal dichalcogenides \cite{lin16, li16bis, Pizzochero17, Pizzochero_2018, Pedramrazi19, Avsar19}.  However, the role of defects in atomically thin magnets remains entirely unexplored to date.

In this Letter, we investigate the formation of intrinsic defects in monolayer CrI$_3$ by means of first-principles calculations.  We address the atomic structure, thermodynamic stability, diffusion and aggregation tendencies along with the defect-induced magnetic properties of a large set of native impurities, including vacancy, adatom, self-interstitial and antisite defects. Overall, this work offers a theoretical insight into defects formation in monolayer CrI$_3$ and further lays the foundation for defect engineering in this prototypical two-dimensional magnet.

\section{II.\ Methodology}
Our first-principles calculations are performed within the spin-polarized density functional theory formalism, as implemented in the the Vienna \emph{Ab Initio} Simulation Package (VASP) \cite{Kresse93, Kresse96}. The exchange and correlation effects are treated under the generalized gradient approximation devised by Perdew, Burke, and Ernzerhof (PBE) \cite{PBE}. In the case of pristine and defective supercells of CrI$_3$, this functional is supplemented with an on-site Coulomb repulsion term $U = 1.5$ eV acting on the $d$ shell of Cr atoms \cite{DFTU}. The resulting PBE+$U$ density functional was shown to provide an excellent description of the magnetic properties of monolayer CrI$_3$, as demonstrated by the comparison with higher-level many-body wavefunction results and available experimental data \cite{Pizzochero19a, Huang2017, Torelli18}.

Electron-core interactions are described through the projector-augmented wave method, while the Kohn-Sham wavefunctions for the valence electrons are expanded in a plane wave basis with a cutoff on the kinetic energy of 400 eV. The integration over the first Brillouin zone is carried out using the equivalent of  $12 \times 12$ $k$-points per unit cell. The atomic coordinates are optimized until the maximum component of the Hellmann-Feynman forces are smaller than 0.02 eV/{\AA}, while constraining the lattice constant to the experimental value of 6.867 {\AA}. The saddle points for the determination of the energy barriers are located with the help of the climbing-image nudged elastic band method \cite{henkel00}. Specifically, we have optimized under the appropriate constraints the intermediate image between the initial and the final states until forces dropped down to 0.04 eV/{\AA}. We model defective lattices by introducing point defects in otherwise pristine $4 \times 4$ hexagonal supercells of monolayer CrI$_3$ containing 128 atoms. A vacuum region 17 {\AA} thick is included to  avoid artificial interactions between periodic replicas.

\section{III.\ Results and Discussion}
  \begin{figure}[]
  \centering
  \includegraphics[width=1\columnwidth]{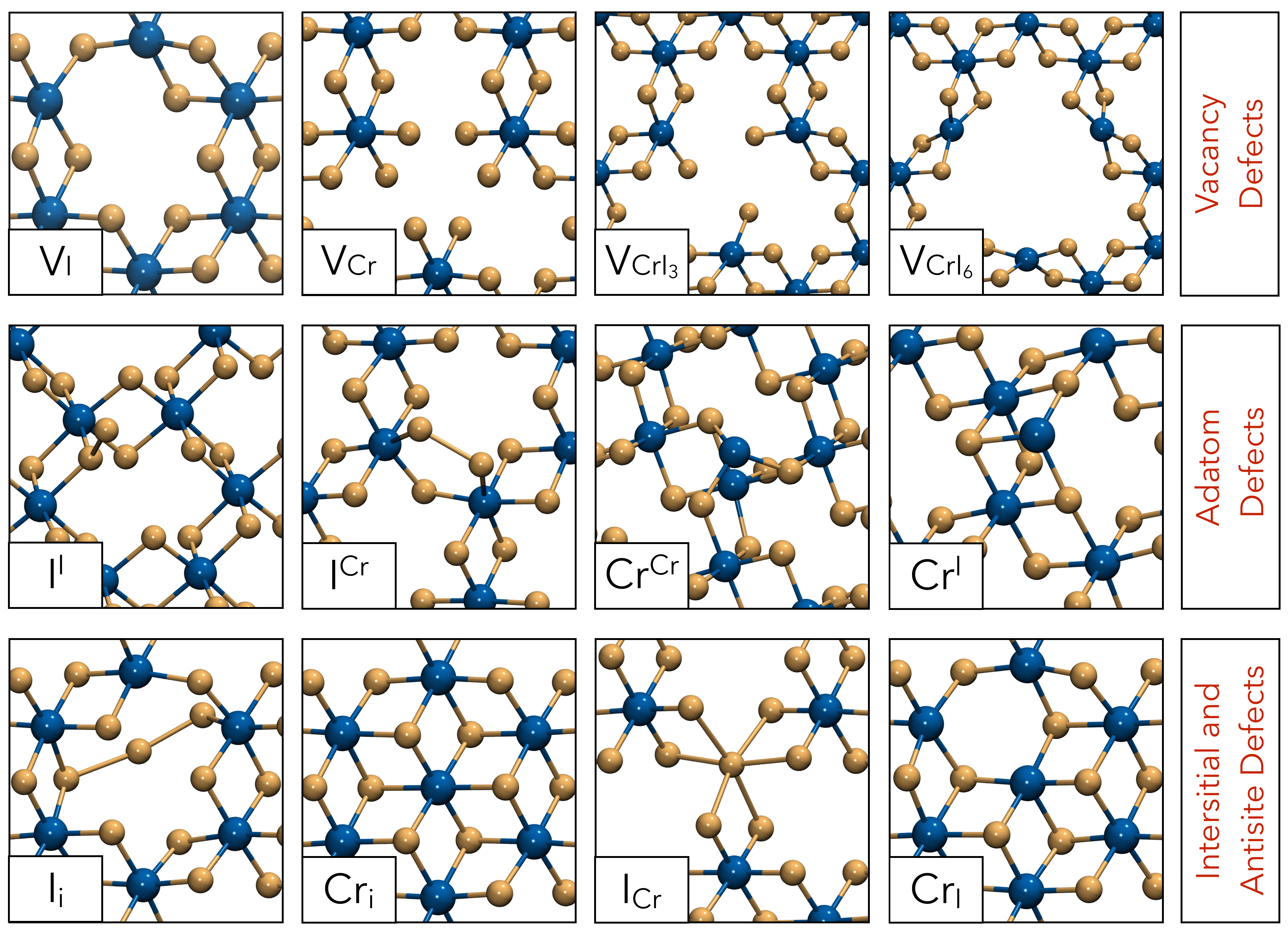}
  \caption{Atomic structure of the point defects in monolayer CrI$_3$ considered in this work. Blue and orange balls represent Cr and I atoms, respectively.} \label{Fig1}
\end{figure}

We start by assessing the stability of native defects in monolayer CrI$_3$. We consider a broad set of different point defects, the atomic structure of which is shown in Fig.\ \ref{Fig1}. The investigated defects belong to three classes, namely (i) vacancy defects, including atomic vacancies (V\textsubscript{I} or V\textsubscript{Cr}) along with their complexes V\textsubscript{CrI\textsubscript{3}} and V\textsubscript{CrI\textsubscript{6}}, (ii) adatom defects, consisting of excess atoms located on top of either chromium (I\textsuperscript{Cr} and Cr\textsuperscript{Cr}) or iodine lattice site  (I\textsuperscript{I} and Cr\textsuperscript{I}), and (iii) antisite (Cr\textsubscript{I} and I\textsubscript{Cr}) or self-interstitial (Cr\textsubscript{i} and I\textsubscript{i}) defects, depending on whether the extra Cr and I atoms are sitting in either an occupied or unoccupied site, respectively.

Under thermodynamic equilibrium, the quantity of primary interest in governing the relative stability of defects is their formation energy ($E\textsubscript{form}$). Given that CrI$_3$  is a binary system, in most cases the introduction of intrinsic defects is accompanied by a deviation from the nominal stoichiometry, thereby rendering $E\textsubscript{form}$ dependent on chemical potentials of the constituent elements. For charge-neutral defects, the formation energy takes the form
\begin{equation}
E\textsubscript{form} (\mu) = E\textsubscript{defect} - E\textsubscript{host} - \sum_i n_i \mu_i
\label{Eq1}
\end{equation}
with $E\textsubscript{defect}$ and $E\textsubscript{host}$ being the total energies of the defective and pristine supercell, respectively, $n_i$ the number of atoms that one has to add to or remove from the lattice in order to create the defect, and $\mu_i$ the corresponding chemical potential. As is customary, we assume that the chemical potential of Cr and I are in thermal equilibrium with that of monololayer CrI$_3$ according to the expression
\begin{equation}
\mu\textsubscript{CrI\textsubscript{3}}  = \mu\textsubscript{Cr} + 3\mu\textsubscript{I}.
\end{equation}
The relevant interval of chemical potential in which defect formation energies are physically meaningful spans the energy range in which monolayer CrI$_3$ remains stable against the decomposition into its elemental constituents. These are taken to be the lowest-energy phases of these chemical elements, namely the bcc Cr crystal and the isolated I$_2$ molecule, with corresponding chemical potentials $\mu\textsubscript{Cr,Cr\textsubscript{bulk}}$ and $\mu\textsubscript{I,I\textsubscript{2}}$, respectively. With these quantities at hand, we are in a position to determine the appropriate chemical potentials of Eqn.\ (\ref{Eq1}). Under I-rich (or Cr-poor) condition, these read as $\mu\textsubscript{I} = \mu\textsubscript{I,I\textsubscript{2}}$ and  $\mu\textsubscript{Cr} = \mu\textsubscript{CrI\textsubscript{3}}  - 3 \mu\textsubscript{Cr,Cr\textsubscript{bulk}}$. On the other hand, under I-poor (or Cr-rich) condition, one finds  $3\mu\textsubscript{I} =\mu\textsubscript{CrI\textsubscript{3}} - \mu\textsubscript{Cr,Cr\textsubscript{bulk}}$  and $\mu\textsubscript{Cr} = \mu\textsubscript{Cr,Cr\textsubscript{bulk}}$. Within this formalism, the range of stability of monolayer CrI$_3$ is $-0.77$ eV $< \mu\textsubscript{I} < 0$ eV.

  \begin{figure}[]
  \centering
  \includegraphics[width=1\columnwidth]{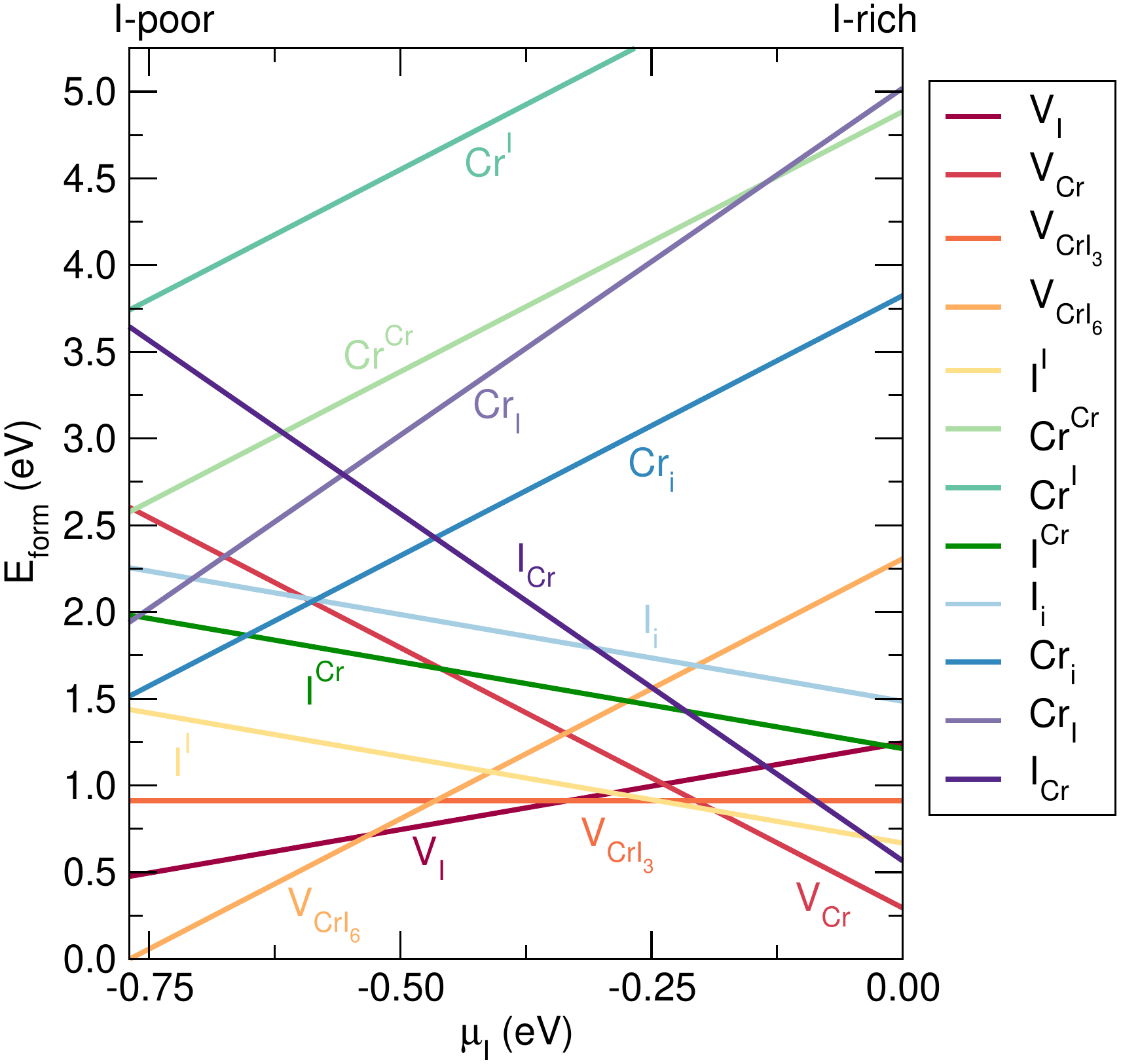}
  \caption{Formation energy ($E\textsubscript{form}$) of the point defects shown in Fig.\ \ref{Fig1} as a function of the iodine chemical potential ($\mu$\textsubscript{I}).} \label{Fig2}
\end{figure}

In Fig.\ \ref{Fig2}, we overview the formation energy of the point defects shown in Fig.\ \ref{Fig1} as a function of the I chemical potential. Although the relative stability of the defects is largely ruled by the chemical potential, our calculations reveal that either atomic vacancies or their complexes exhibit the lowest formation energies. Specifically, I-rich conditions promote the formation of V\textsubscript{Cr} and V\textsubscript{CrI\textsubscript{3}} defects, whereas I-poor conditions enhance the stability of V\textsubscript{I} and V\textsubscript{CrI\textsubscript{6}} ones. This result demonstrates that, when thermodynamic equilibrium prevails, monolayer CrI$_3$ is likely to exhibit a sub-stoichiometric composition.

As far as the metastable defects are concerned, we found that I adatoms display lower formation energies than Cr adatoms over the entire range of chemical potential. However, while the binding of an extra I atom on top of an I site is more stable than on top of a Cr site by 0.55 eV, the opposite behavior is observed for an extra Cr atom, being Cr\textsuperscript{Cr} lower in energy than Cr\textsuperscript{I} by 1.15  eV. Differently from adatom defects, the relative stability of self-interstitials is dominated by the chemical potential conditions. In fact, the I\textsubscript{i} defect remains the most stable self-interstitial defect only under I-rich conditions, as approaching I-poor conditions stabilizes the formation of the Cr\textsubscript{i} defect. It is worth noticing that, while excess I atoms incorporate into monolayer CrI$_3$ as adatom defects (\emph{i.e.}, I\textsuperscript{I} is more stable than I\textsubscript{i} by 0.82 eV), excess Cr atoms should instead be accommodated in the lattice as self-interstitials (\emph{i.e.}, Cr\textsubscript{i} is more stable than Cr\textsuperscript{Cr} by 1.06 eV). The relative stability of antisites is also governed by chemical potential conditions, as I-rich (I-poor) conditions stabilize the I\textsubscript{Cr} (Cr\textsubscript{I}) defect. We notice that the formation energy of both the antisite defects considered is larger than the sum of the formation energy of an isolated vacancy and an isolated adatom by 0.40 eV and 0.05 eV for I\textsubscript{Cr} and Cr\textsubscript{I} defects, respectively. This indicates that vacant sites enhance the reactivity of the lattice as compared to the pristine areas of the monolayer.

  \begin{figure}[]
  \centering
  \includegraphics[width=1\columnwidth]{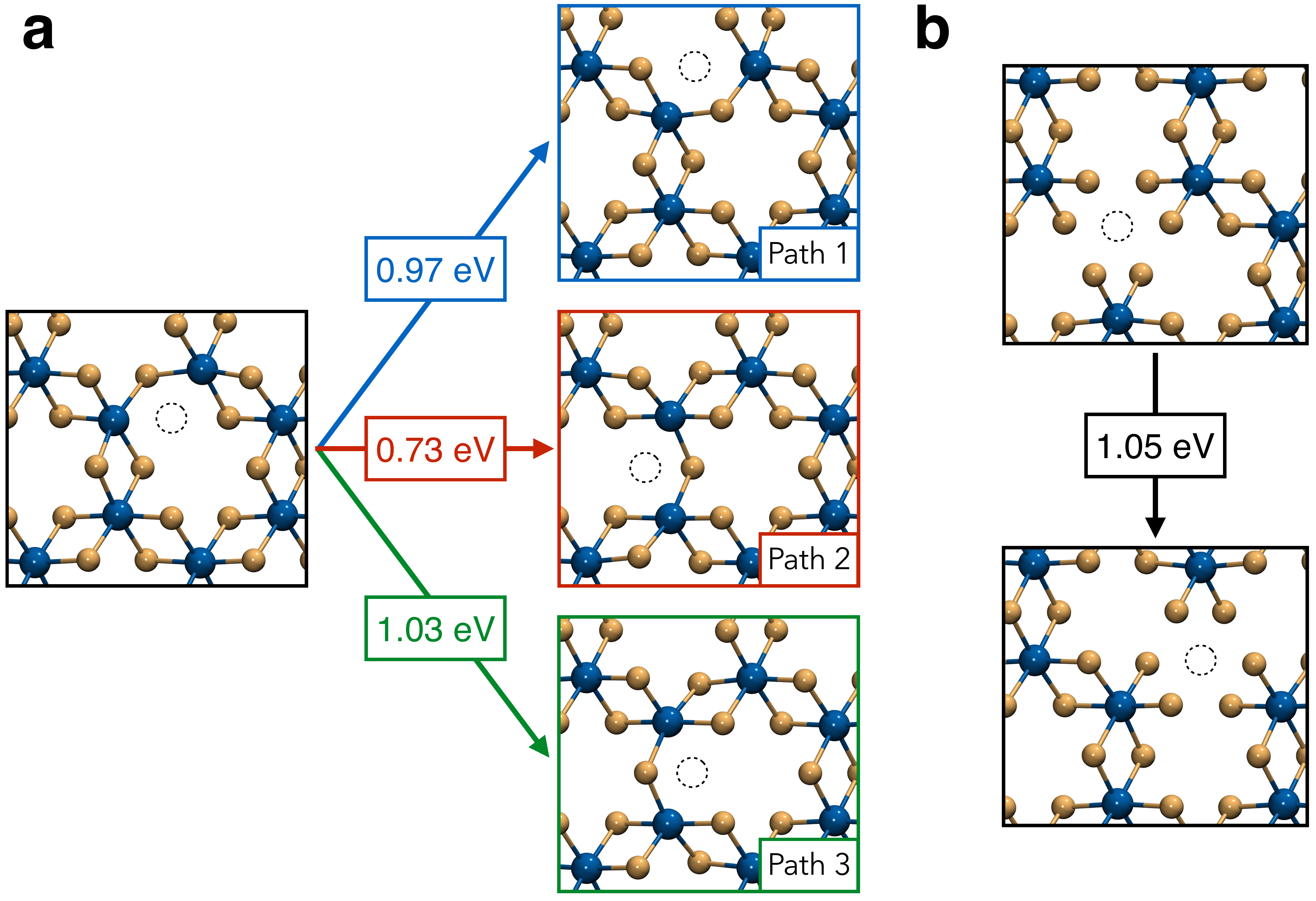}
  \caption{Energy barriers to diffuse to neighboring sites for (a) I and (b) Cr vacancy defects. The defect site is marked with a black dashed circle.} \label{Fig3}
\end{figure}

We now turn our attention to the mobility of atomic vacancies. As we discussed above, these defects exhibit the lowest formation energy in a quite broad range of chemical potential and further constitutes the building blocks for the also stable vacancy complexes V\textsubscript{CrI\textsubscript{3}} and V\textsubscript{CrI\textsubscript{6}}. In Fig.\ \ref{Fig3}(a), we give the energy barriers for a V\textsubscript{I} to diffuse to the three nearest neighboring I sites. Specifically, we consider diffusion processes to either the I atom bridging the same (path 1) or the nearest (paths 2 and 3) pair of Cr atoms between which V\textsubscript{I}  is introduced. We found that V\textsubscript{I} features a diffusion barrier in the 0.73 -- 1.03 eV range, depending on the path considered, with path 2 being the most energetically favorable. Furthermore, we determine the energy barrier for a V\textsubscript{Cr} defect to diffuse to the nearest equivalent to be 1.05 eV, as we show in Fig.\ \ref{Fig3}(b). In order to estimate the diffusion rate $r$ of vacancy defects, we rely on the the Eyring equation \cite{Eyring}. Assuming a negligible change in entropy between the initial and transition states, this reduces to

\begin{equation}
r = \frac{k\textsubscript{B} T}{h} \exp \left[ -\frac{E\textsubscript{barr}}{k\textsubscript{B}T}\right]
\end{equation}
with ${E\textsubscript{barr}}$ being the energy barrier for vacancy defects to diffuse (see Fig.\ \ref{Fig3}), $T$ the considered temperature, $k$\textsubscript{B} and $h$ the Boltzmann and Planck constants. At  $T = 300$ K, we find  $r \approx 10^{1}$ s$^{1}$ and $r \approx 10^{-4}$ s$^{-1}$ for V\textsubscript{I} (along the favorable path 2) and V\textsubscript{Cr}, respectively, hence indicating a certain degree of mobility of these defects at room temperature.

  \begin{figure}[]
  \centering
  \includegraphics[width=0.85\columnwidth]{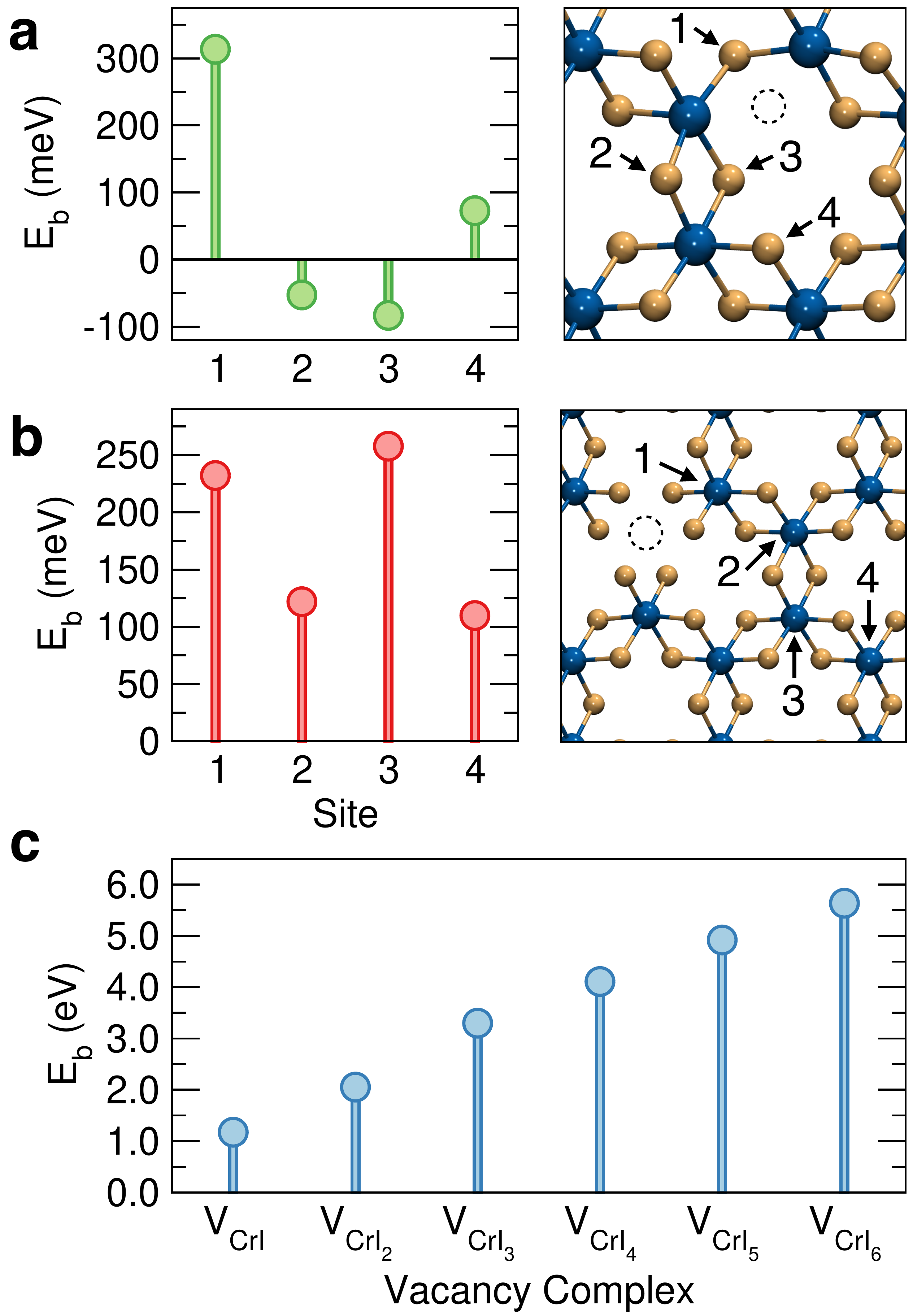}
  \caption{Agglomeration energy ($E\textsubscript{b}$) of a pair of (a) I and (b) Cr vacancy  defects. The site at which the first vacancy is located is indicated with a black dashed circle. (c) Agglomeration energy of vacancy complexes V\textsubscript{CrI\textsubscript{x}}, with $x = 1, 2, ..6$.} 
  \label{Fig4}
\end{figure}

Next, we investigate the tendency of vacancy defects to aggregate. To this end, we quantify the agglomeration energy $E\textsubscript{b}$ of a given defect complex $AB$, formed upon clustering of the otherwise spatially separated $A$ and $B$ defects, by comparing their formation energies as
\begin{equation}
E\textsubscript{b} =  E\textsubscript{form}(A) + E\textsubscript{form}(B) - E\textsubscript{form}(AB).
\end{equation}
According to this expression, positive (negative) values of $E\textsubscript{b}$ point towards an energetically favorable (unfavorable) tendency of defects to aggregate. In Fig.\ \ref{Fig4}(a,b), we show the agglomeration energies to form a secondary vacancy defect at several lattice sites neighboring the first missing atom. For the I divacancies case shown in Fig.\  \ref{Fig4}(a), we find that, depending on the site considered, the introduction of a secondary V\textsubscript{I} can be either energetically favorable (sites 1 and 4) or unfavorable (sites 2 and 3), with the agglomeration energy attaining its maximum upon formation of a secondary I vacancy at site 1. This is in contrast with the formation of a second V\textsubscript{Cr} shown in Fig.\ \ref{Fig4}(b), whereby  $E\textsubscript{b}$ remains positive at all sites investigated. In particular, we observe that the formation of a Cr divacancy defect is more stable when involving two Cr atoms bridged by the same pair of I ligands (that is, sites 1 and 3).

In order to understand the stability of the vacancy complexes V\textsubscript{CrI\textsubscript{3}} and V\textsubscript{CrI\textsubscript{6}}, we calculate their agglomeration energy with respect to isolated Cr and I vacancies. Specifically, we remove an increasing amount of I atoms residing in the first coordination shell of a V\textsubscript{Cr} defect, therefore giving rise to a family of V\textsubscript{CrI\textsubscript{x}} defects, with $x = 1, 2,..., 6$. Our results are given in Fig.\ \ref{Fig4}(c). For all considered V\textsubscript{CrI\textsubscript{x}} complexes, we find a substantial aggregation tendency, with $E\textsubscript{b}$ in the of 1--6 eV energy interval and monotonically increasing by $\sim$1 eV for each subsequent I atom removed. These values are at least one order of magnitude larger than those found for I and Cr divacancies. We suggest that such a marked tendency of I vacancies to agglomerate around a V\textsubscript{Cr} defect is at the origin of the thermodynamic stability of V\textsubscript{CrI\textsubscript{3}} and V\textsubscript{CrI\textsubscript{6}} observed in Fig.\ \ref{Fig2}.

\begin{table}[]
\caption{%
 {Difference in local magnetic moments between pristine and defective monolayers CrI$_3$ ($\Delta \mu$), the latter hosting the point defects shown in Fig.\ \ref{Fig1}.}
}
\begin{ruledtabular}
\begin{tabular}{ lclc}
\textrm{Point Defect}& \textrm{$\Delta \mu$ ($\mu\textsubscript{B}$)} & \textrm{Point Defect}& \textrm{$\Delta \mu$ ($\mu\textsubscript{B}$)} \\
\colrule
V\textsubscript{I} & $1.00$ & Cr\textsuperscript{Cr} & $6.00$ \\
V\textsubscript{Cr} & $-6.00$ & Cr\textsuperscript{I} & $6.00$  \\
V\textsubscript{CrI\textsubscript{3}}  & $-3.00$ &Cr\textsubscript{i} & $1.00$   \\
V\textsubscript{CrI\textsubscript{6}}& $0.00$ & I\textsubscript{i} & $6.00$ \\
I\textsuperscript{I} & $1.00$ & I\textsubscript{Cr} & $-5.00$ \\
I\textsuperscript{Cr} & $1.00$ & Cr\textsubscript{I} & $7.00$
\label{Tab1}
\end{tabular}
\end{ruledtabular}
\end{table}

Finally, we discuss the effect of point defects on the magnetism of CrI$_3$. Table \ref{Tab1} lists the difference in magnetic moments between pristine (\emph{i.e.}, 3 $\mu\textsubscript{B}$ per Cr$^{3+}$ ion, in accord with experiments \cite{Huang2017}) and defective lattice, this latter hosting the defects shown in Fig.\ \ref{Fig1}. Remarkably, we notice that all defects but V\textsubscript{CrI\textsubscript{6}}  affect the magnetism of monolayer CrI$_3$, thereby suggesting that deviations from the nominal stoichiometry are likely to be accompanied by an alteration of the local magnetic moments. Specifically, V\textsubscript{I} and  V\textsubscript{Cr}  defects lead to a $\Delta \mu$ of $-1.00$ $\mu\textsubscript{B}$ and $6.00$ $\mu\textsubscript{B}$, respectively, while the $\Delta \mu$ associated with the V\textsubscript{CrI\textsubscript{3}} complex is  $-3.00$ $\mu\textsubscript{B}$. Excess Cr or I atoms, whether as adatom or self-interstitial defects, change the intrinsic magnetic moments by $6.00$ $\mu_B$ and $1.00$ $\mu_B$, respectively, while antisite I\textsubscript{Cr} (Cr\textsubscript{I}) defect induces a larger $\Delta \mu$ of $-5.00$ $\mu_B$ ($7.00$ $\mu\textsubscript{B}$).

\section{IV.\ Conclusion}
In summary, we have carried out extensive first-principles calculations to investigate the formation of native defects in the prototypical two-dimensional magnet CrI$_3$.  Depending on the chemical potential conditions, we have identified mobile V\textsubscript{I} and  V\textsubscript{Cr} species along with their complexes V\textsubscript{CrI\textsubscript{3}} and V\textsubscript{CrI\textsubscript{6}} as the most stable defects, these latter emerging as a consequence of the strong agglomeration tendency. Our results indicate that, under thermodynamic equilibrium, monolayer CrI$_3$ should exhibit a sub-stoichiometric nature, which in turn is found to affect the magnetic moments locally. To conclude, our findings pinpoint a marked intertwining between atomic-scale disorder and magnetism in monolayer CrI$_3$.

\section{Acknowledgments}
I wish to gratefully acknowledge Oleg V.\ Yazyev, Kristi\={a}ns \v{C}er\c{n}evi\v{c}s, Nikita Tepliakov and Ravi Yadav at EPFL for fruitful interactions.
I am financially supported by the Swiss National Science Foundation (SNSF) through the Grant No.\  172543. First-principles calculations were performed at the Swiss National Supercomputing Centre (CSCS) under the project s832.

 \bibliography{References}

 \end{document}